# Modelling the impact of Multi Cancer Early Detection tests: a review of natural history of disease models

**Mandrik O[1], Whyte S[1], Kunst N[2], Rayner A[1], Harden M[3], Dias S[3], Payne K[4], Palmer S[2], Soares MO[2],***

1  Sheffield Centre for Health and Related Research, University of Sheffield
2  Centre for Health Economics, University of York
3 Centre for Reviews and Dissemination, University of York
4 Manchester Centre for Health Economics, School of Health Sciences, The University of Manchester, UK

* Corresponding author: Soares MO

**Funding statement**: Financial support for this study was provided by NHS England. The funding agreement ensured the authors' independence in designing the study, interpreting the data, writing, and publishing the report.

**Author contributions statement**: Harden designed the literature search strategy; Kunst conducted the search; Mandrik, Rayner and Whyte extracted and summarised the models; Soares and Palmer conducted the critical appraisal; Soares drafted the article; all authors contributed to the write-up and approved the final version.

**Declaration of conflicting interests**: The Authors declare no potential conflicts of interest with respect to the research, authorship, and/or publication of this article.

**Acknowledgement**: We thank GRAIL for providing access to their models and for related discussions.

# Abstract

Introduction: The potential for multi-cancer early detection (MCED) tests to detect cancer at earlier stages is currently being evaluated in screening clinical trials. Once trial evidence becomes available, modelling will be necessary to predict impacts on final outcomes (benefits and harms), account for heterogeneity in determining clinical and cost-effectiveness, and explore alternative screening programme specifications. The natural history of disease (NHD) component of a MCED model will use statistical, mathematical or calibration methods.

Methods: Modelling approaches for MCED screening that include an NHD component were identified from the literature, reviewed and critically appraised. Purposively selected (non-MCED) cancer screening models were also reviewed. The appraisal focussed on the scope, data sources, evaluation approaches and the structure and parameterisation of the models.

Results: Five different MCED models incorporating an NHD component were identified and reviewed, alongside four additional (non-MCED) models. The critical appraisal highlighted several features of this literature. In the absence of trial evidence, MCED effects are based on predictions derived from test accuracy. These predictions rely on simplifying assumptions with unknown impacts, such as the stage-shift assumption used to estimate mortality impacts from predicted stage-shifts. None of the MCED models fully characterised uncertainty in the NHD or examined uncertainty in the stage-shift assumption.

Conclusion: MCED technologies are developing rapidly, and large and costly clinical studies are being designed and implemented across the globe. Currently there is no modelling approach that can integrate clinical study evidence and therefore, in support of policy, it is important that similar efforts are made in the development of MCED models that make best use of the available data on benefits and harms.

# 1 Introduction

Novel technologies have recently emerged that look for markers of cancer in blood, urine, saliva or stool and have the potential to detect signals from multiple cancer types from a single sample. These are termed multi-cancer early detection (MCED) tests. Their use in screening asymptomatic persons has the potential to detect cancer at an earlier stage, when treatment is likely to be more effective and perhaps less costly.[1, 2] However, policymakers have demanded evidence of mortality impacts, and a fuller examination of the potential harms and consequences of the test's imperfect accuracy (including of diagnostic resolution pathways), of overdiagnosis, and of the impact on existing screening programmes.[3] The Galleri® test (GRAIL, Inc., Menlo Park, CA, USA) test is the blood multi-cancer test that is most advanced in the stage of clinical research, with a randomised clinical trial currently underway in the UK, the NHS-Galleri trial (NCT05611632), aiming to demonstrate the clinical effectiveness of the test in stage-shifting advanced cancer in a population screening setting.[4]

To inform policy decisions on screening programmes involving MCED tests, modelling will be required to i) link evidence and predict expected impacts over final outcomes (mortality, life expectancy and Quality Adjusted Life Years, or QALYs), ii) appropriately reflect heterogeneity in the value of stage shifts across different cancer types to allow estimation of cost-effectiveness, and iii) allow alternative specifications for a screening programme to be evaluated (e.g. different age and risk groups, alternative screening intervals, etc.). Modelling is therefore likely to underpin such policy evaluations of MCED tests. This may include statistical, mathematical or calibration modelling to integrate cancer screening data and infer the natural history of disease. It may also include decision modelling to predict results with alternative screening regimens and their longer term clinical and cost-effectiveness.

Cancer screening models typically include a natural history of disease (NHD) component that describes the prevalence of preclinical cancer (undiagnosed but detectable) and allows for examining the impact of important policy options, such as alternative specifications for the screening program. The NHD model component describes cancer progression through its preclinical stages over time (in the absence of the proposed screening test) and may also consider cancer onset and the competing risks of clinical detection (both incidental findings and symptomatic presentation) and mortality. The challenge in evaluating these NHD models arises from the fact that preclinical progression is unobserved. Empirical data, however, can still provide relevant information on preclinical cancer prevalence and progression supporting inference –where the data are used to infer the NHD model and help gain an understanding of the likely values of the NHD model parameters in the underlying population, using statistical and mathematical approaches[5] or calibration[6, 7]. Besides alternative evaluation approaches, models in the general cancer screening literature[8, 9] also use a variety of data sources and analytical methodologies, vary the core elements of the NHD that are modelled (they may or may not model cancer onset, the likelihood of clinical detection and/or mortality), and vary whether and how within-tumour heterogeneity and overdiagnosis are modelled.

The objective of this paper is to identify, review and critically appraise the existing literature for alternative modelling approaches proposed for MCED that include an NHD component. As the literature and approaches in this area continue to develop and evolve, it is important to critically examine the range of modelling approaches that have been proposed for MCEDs, and to assess the extent to which specific features of model structure and model evaluation can accommodate the complexity of multi-cancer modelling. While there has been extensive discussion and consideration of the appropriate study design to inform clinical utility[10], we are not aware of any publications that have attempted to systematically identify and critique existing modelling approaches and specifically the extent to which they will be able appropriately integrate the findings of these clinical utility studies. The paper is structured in the following way: the existing models are identified and described in Section 2, critically appraised in Section 3 and overall findings are discussed in Section 4. A glossary (Box 1) of definitions will be used throughout.

<<insert Box 1 here>>

## 2 Review of models

### 2.1 Methods

#### 2.1.1 Literature search for MCED models

A scoping literature review was developed and undertaken to identify published models of MCEDs in relation to a comparator. The review included models of NHD that incorporate both detection rates and predicted stage distribution (stage-shift), and which may also have extended these models to quantify impacts on mortality. The search methodology is reported in full in the Online Appendix.

#### 2.1.2 Additional selected models

To further support the critical appraisal, additional selected models, including multi-disease (but non-MCED) models and single disease models cited by relevant authors to support existing MCED models, were added purposively. For these additional models, the focus of the review was on the modelling mechanisms and related assumptions. These models provide background and context for the modelling assumptions made in MCED models, and for any changes/extensions made.

## 2.1.3 *Extraction*

The review extracted information on the following aspects: i) model structure, including the number and types of cancer, NHD parameterisation and modelling of screening impact, ii) how data were used within the models, including key NHD assumptions, data requirements, and iii) uncertainties related to NHD and how these were considered. From the extracted information, we identified the following features of the models reviewed:

- Scope: The population modelled (whether only individuals with clinically diagnosed cancer were modelled or the entire population eligible for screening, which would have allowed quantifications of overdiagnosis) and whether mortality impacts were considered.

- Key data sources: Whether evidence on detection with screening was considered (e.g. clinical trial) or only evidence on cancer incidence under current care; whether external evidence on pre-clinical progression parameters (elicited or from the literature) was used.

- Evaluation: Whether the model evaluation was based on prediction or on inference, and whether it is evaluated at the cohort or individual level.

  A predictive approach uses input evidence to directly describe model parameters and calculates expected cancer detection algebraically, with and without screening. NHD model parameters are pre-specified using values or distributions (using external sources like other evaluations or expert opinion), before running the model, which then outputs predictions. In contrast, inferential approaches use cancer diagnosis data from samples of individuals (for example, repeat screening data) to learn about NHD model parameters. Because sojourn time is not directly observable, the methods used differ from standard regressions and employ mathematical techniques such as deconvolution [11] or calibration [6, 7, 12].

- Structure and parameterisation: whether a common structure across cancer types is used; what level of disaggregation of cancer stages was used (i.e. whether individual stages were considered or whether they were aggregated e.g. early vs. late cancer); whether the impact of screening is predicted from test accuracy; whether mortality impact is predicted by applying mortality in clinically detected cancer to the screening stage-distribution predicted by the model; what parameterisation, distributional assumptions, and assumptions about correlation between progression parameters were used; whether overdiagnosis (definition in Box 1) is quantified within the NHD model.

## 2.2 Results

The review identified five different MCED models with an NHD component: four funded by GRAIL (hereafter termed 'GRAIL models') and specifically related to the Galleri test [13-16] and one based on a hypothetical MCED (although using some inputs derived from the Galleri test) [17]. Four additional models (non-MCED) were also reviewed: two multi-cancer models by Thomas (communication with author) and Mandrik [18], and two single-cancer models by Pinsky[11] and Skates [19]. The nine models were reviewed; these are described in Table 1.

<<insert Table 1 here>>

### 2.2.1 MCED models

Table 2 describes the key features of the models reviewed. These models are referred to by the name of the first author in the publication.

<<insert Table 2 here>>

There are four GRAIL models: Hubbell, Sasieni, Tafazzoli and Dai [13-16]. These use a common approach, referred to as the 'interception model', to determine the NHD and stage-shift with the Galleri test, with the core methodology rooted in the Hubbell model. These also use a common set of evidence, including national cancer incidence statistics (by type, stage, age, and gender), expert or literature-derived pre-clinical progression evidence [13, 20] and test sensitivity from diagnostic studies.

The NHD component of the GRAIL models focuses on individuals clinically diagnosed with cancer under standard care. The GRAIL models use a common NHD structure, assuming (Table 3): i) disease progression across four stages (stages 1, 2, 3, and 4) without regression, ii) progression is sequential, with cancers moving through each stage until clinically detected, iii) sojourn times are exponentially distributed, iv) sojourn times are independent between stages, v) there is no heterogeneity in sojourn times within-tumour types beyond that expected by chance (i.e. the expected value of the sojourn time is equal for all individuals in the model). The NHD does not include the probability of cancer onset nor of clinical detection.

<< insert Table 3 here>>

All four GRAIL models consider stage-shift as the main clinical benefit of screening. Stage-shift is evaluated predictively, using test sensitivity to determine the likelihood of earlier detection. Mortality effects are also predicted under the 'stage-shift assumption' and the 'lead time assumption' (Box 1), except in a scenario of the Tafazzoli model that considers mortality during lead time.

None of the GRAIL models reviewed evaluated uncertainty probabilistically. In the context of predictive modelling, this would have entailed describing uncertainty in the input parameters and running probabilistic analysis to evaluate uncertainty over models' outputs. Also, models only incorporate within-tumour heterogeneity from the distribution of cancer diagnosis by age and sex. Further consideration of these aspects are provided in the critical appraisal and discussion sections.

Although the GRAIL models are underpinned by the same core methodology proposed in Hubbell, there are a number of specific differences in terms of their parameterisation and structural assumptions. The Sasieni model [15, 21] applies the Hubbell model to UK cancer incidence and mortality data and examines structural extensions allowing consideration of differential survival of cfDNA-detectable cancers, alternative cohorts and screening regimens, and the possibility of non-sequential progression from stage I to IV only. Tafazzoli's model [16] integrates Hubbell's stage-shift matrices (i.e. the likelihood of a cancer clinically detected in a particular stage being detected by Galleri at each earlier cancer stage) within a cohort model of 50 year-old individuals tested annually with Galleri until the age of 79 years. In Tafazzoli's model, stage shifted individuals in each model cycle are time-shifted (shifted back in time to earlier cycles to account for an earlier time of diagnosis), based on cancer-specific sojourn times. Tafazzoli is the only Galleri model which incorporates overdiagnosis (but not explicitly in the NHD model) by increasing detection by a proportion that is applied as an input to the model, and extends the evaluation to cost-effectiveness. Dai's model [13] uses the core assumptions of Hubbell's model but evaluates the model using individual patient simulation. It also describes sojourn times from empirically derived estimates sourced from other screening studies, rather than elicitation.

Our review identified only one MCED model that was not funded by GRAIL: this is Lange's model [17]. This model examines the impact of a hypothetical MCED (using the estimates for test sensitivity that are relevant to Galleri) on 12 cancer types. The model does not evaluate overdiagnosis or mortality (extensions to mortality have been further considered since publication, see https://cedarmodelingframework.shinyapps.io/mcedmodel/). It is based on the same type of evidence as the GRAIL models (age and stage-specific clinical incidence data under current care) but applies an alternative NHD model that is more comprehensive in that it, in addition to pre-clinical progression (for which it uses the more aggregate classification of early vs. late disease), also characterises probability of cancer onset, the age of cancer onset, and the likelihood of clinical detection. Lange evaluates the underlying NHD model parameters using an inferential approach to describe clinical incidence rate data using a Poisson distribution; however, not all parameters of the model are identifiable based on age and stage-specific incidence data. Therefore, given these data, the authors assumed fixed values for overall and late-stage sojourn times (user-defined) allowing estimating all unknown parameters. It is unclear how inference over the early-stage sojourn times is, however, reached.

### 2.2.2 Selected non-MCED models

*Selected multi-cancer screening models*

Mandrik's model [18] examines the clinical- and cost-effectiveness of a urine dipstick test in screening for bladder and kidney cancers. The NHD model structure includes cancer onset, preclinical cancer progression through cancer stages (1 to 4), cancer detection and mortality. Heterogeneity is included by considering cancer onset to depend on age and smoking status and by considering a separate cancer pathway for non-fatal low risk bladder cancers. Mandrik's model uses detection data for current care only (due to the absence of data for the screening test under evaluation) and summary evidence from the literature on the impact of risk factors, test sensitivity and other elements. The model is Markovian for all transitions except for progression of pre-clinical cancers which uses an individual patient time-to-event formulation. The NHD model was evaluated using Bayesian calibration (Metropolis-Hastings algorithm), an inferential calibration procedure which allows for uncertainty to be appropriately integrated. Due to the absence of screening data, and to ensure model identification, strong priors, assumptions and constraints over the NHD parameters were used. A predictive approach anchored on test accuracy was used to project screening outcomes from test accuracy, overdiagnosis and mortality impacts (from stage-shifts)

Thomas' model (unpublished) evaluates upper abdominal CT imaging for the screening of ten cancers (alongside other abdominal diseases). It adopts a common structure across all cancers, with progression across stages 1 to 4. The model uses clinical incidence data with screening, combined with elicited estimates of test sensitivity. Despite not considering probability of cancer onset, the model considers the age of onset in those that were screen-detected. For the comparator arm, the model simulates what would have happened to the screen-detected individuals had they not been screened. In doing so, it considers the competing events of stage progression, clinical detection and mortality in its structure. The model is a multi-cohort Markov model, considering various age and sex cohorts. The model conducts inference using a simplified non-Bayesian calibration (or fitting process), which does not consider uncertainty over the NHD, to evaluate outcomes for a cohort of unscreened individuals from elicited values describing stage-specific pre-clinical progression. Mortality impacts were predicted from stage-shifts. By considering that those that would have been screen-detected were at risk of death if unscreened, the comparator arm considers individuals dying with undiagnosed cancer and predicts a lower number of cancer cases than in the screening arm.

*Single-cancer models cited by authors of existing MCED models*

Skates' model [19] is cited in the GRAIL models in support of the proposed interception model. Skates examines the impact of screening for ovarian cancer with a blood biomarker using a predictive approach combining ovarian cancer incidence with pre-clinical progression times across 4 cancer stages. The key difference between this NHD model and the GRAIL NHD models are that Skates uses patient-level simulation (all GRAIL models except for Dai), a different parameterisation of time to stage progression using log-normal distributions with fixed mean ratios between stages and a coefficient of variation, and accounts for correlation between stages. The impact of screening is predicted from biomarker levels, and the mortality impact is predicted using the 'stage-shift assumption' and the 'lead time assumption' while also assuming a proportion of patients are cured.

Pinsky's model [11] is the key reference cited by Lange. It uses the same structure as Lange but considers a range of distributions for the NHD parameters and imposes age dependency on time to cancer onset. Pinsky's model, however, uses screening trial data to achieve inference on the NHD via maximum likelihood estimation. In doing so, it carefully considers parameter identifiability from the data.

## 3    Critical appraisal

In this section, we critically appraise the existing MCED models for their key features, including how these accommodate the multi-cancer context, and highlight key uncertainties.

*MCED effects are based on predictions rather than direct evidence.*

A critical feature of this evaluation problem is the current absence of data on cancer detection and mortality from screening with Galleri or other MCED. The NHD models therefore use similar data, namely on cancer incidence data under current care and expected sojourn times, to back-calculate or infer undiagnosed cancer prevalence. The lack of screening data means that the accuracy of predictions and inferences in MCED models will rely on the use of an appropriate NHD model and on the quality of the evidence underlying/supporting the NHD parameters.

*MCED models apply simplifying assumptions. It is unclear where adding complexity may be most important.*

Existing MCED models, despite the similarity in the data included, have proposed a wide variety of modelling approaches for the NHD—from predictive to inferential models, cohort to individual level simulation, more complex (or simpler) assumptions over the NHD. MCED models apply the most assumptions (see Table 3) despite many being shared with other models. This may be motivated by the multicancer context and the need to reduce parameterisation and employ simpler evaluation approaches. There has been limited exploration of the impact of these simplifications, and it is unclear where additional complexity may add value.

Some of the simplifying assumptions allow the NHD to be evaluated algebraically as with the GRAIL models (i.e. exponentially distributed pre-clinical progression times with a common mean and independent across stages). However, models run as individual patient level simulation, such as Skates and Mandrik, allow relaxing these assumptions and varying the level of variation (heterogeneity) in sojourn time, i.e. the proportion of cases with extreme sojourn times. Existing explorations are insufficient to identify the likely sources and key impacts of heterogeneity, but suggest important impacts (see, for example, Sasieni's scenario considering a proportion of very fast progressing cancers).

*Overdiagnosis is not explicitly modelled in MCED models and adding this may add complexity to modelling.*

One important potential harm of screening is overdiagnosis. Overdiagnosis has the potential to be explicitly estimated/predicted within an NHD model with a fuller structure that characterises heterogeneity and includes cancer onset and mortality alongside preclinical progression and clinical detection. None of the MCED models have estimated/predicted overdiagnosis within the NHD model, presumably because of the reliance on a restricted structure and scope to allow evaluation from cancer incidence data, e.g. Hubbell only characterised cancer progression and Lange also included cancer onset but not mortality. Of the broader models reviewed, those including a full NHD structure, such as Mandrik, included overdiagnosis, but none explicitly examined whether and how heterogeneity may affect overdiagnosis estimates.

*Current MCED models do not appropriately characterise uncertainty in the NHD*

Decisions in health are often made under uncertainty, and explicit descriptions of uncertainty help determine appropriate funding and research decisions. Uncertainty in model inputs can be described and propagated in prediction modelling, however, none of the predictive MCED models reviewed have done so. Since our review was conducted GRAIL published an extension of the Tafazzoli model that includes probabilistic analysis [22], although, in this analysis, none of the NHD parameters were assumed uncertain (e.g. sojourn times, mortality). Of the MCED models, only Lange considers uncertainty in the NHD by implementing an inferential procedure describing the cancer incidence data as uncertain. However, other important sources of uncertainty were not formally included in Lange's model, such as uncertainty over sojourn times, but can be examined by varying the choice of sojourn time inputs.

*All models predict mortality impacts using the stage-shift assumption*

The 'stage-shift assumption' is only plausible if cancers detected by a screening test do not differ systematically in their characteristics from clinically detected cancers. For example, if the higher ctDNA shedding expected in cancers detected by Galleri is associated with worse prognosis, the capacity of stage-shifted cancers to benefit may be smaller than expected. The Sasieni model examined hypothetical reductions in the capacity to benefit of stage-shifted cancers and showed that impact can be significant. A number of publications have explored the accumulation of evidence, across screening trials, in support of the stage-shift assumption [13, 23, 24]. However, the validity of this assumption for particular multi-cancer tests is unknown until well designed clinical research reports on the mortality impacts. The NHS-Galleri trial, at the time its primary endpoint reports, may not provide sufficient mortality evidence and this is therefore likely to remain a key uncertainty for decision making.

# 4 Discussion

We identified, summarised and critically appraised the NHD components of models of the clinical and/or economic impact of using MCED tests in a screening programme. We have systematised and categorised elements of the modelling approach, evaluation strategy and structure in a way that is novel and relevant for the broader screening modelling literature. We found that MCED models are characterised by the absence of screening data, by the limited use of inference and by the limited characterisation of uncertainty, heterogeneity and overdiagnosis within the NHD. Our critical appraisal identified limitations of current MCED models and highlighted the limited exploration of the impact of modelling assumptions.

Our findings have important implications for future models of the clinical and cost-effectiveness of MCED screening programmes, which will need to incorporate clinical utility study evidence in support of decision making. This requires an inferential approach but, to date, no such approach has been developed to include screening data in the multi-cancer context. There is an extensive literature on inferential approaches used in the single disease context, which include: a) mathematical/statistical models that typically using a single main source of evidence and a clear specification of the model (NHD) with lower dimensionality (e.g. typically aggregating cancer stages for example), and b) calibration models, typically using multiple sources of evidence (as calibration targets) and, perhaps for this reason, a higher dimensionality. In this paper we did not review this broader literature, but the future development of an inferential approach for MCEDs should draw on it.

MCED trials, like the NHS-Galleri trial, are likely to be powered on stage-shift outcomes aggregated over multiple cancer types and estimates for each cancer type will need to be strengthened using modelling alongside additional external evidence. Model identifiability will need to carefully consider higher parameterisations (e.g. more detailed descriptions of between- and within-tumour heterogeneity) and the support of the evidence for structural simplifications in such descriptions and in the potential aggregation across cancer stages. GRAIL models disaggregate across the four cancer stages, but most mathematical approaches aggregate stages into early and advanced cancer or simply distinguish preclinical from clinical cancers. Uncertainty over the 'stage-shift assumption' needs to be examined in further work in support of decision making.

Computational burden is also of concern, as more complex models may compromise transparency and accessibility, particularly for calibration approaches, typically using individual level simulation, applied in the multicancer context. Alternatives to individual level simulation can be considered, such as the multi-cohort model structure exemplified in Thomas. It partitions the cohort into sub-cohorts based on relevant baseline characteristics, such as risk or demographic groups.

Other key considerations for future MCED model development relate to overdiagnosis and within-tumour heterogeneity. In what concerns overdiagnosis, there are important challenges in obtaining valid empirical estimates [25] and therefore decision making may initially need to consider estimates from modelling which require extensions to existing MCED modelling approaches (see critical appraisal section). In what concerns within-tumour heterogeneity, this is known to exist across several cancer types. Heterogeneity has been considered in the broader screening modelling literature structurally, for example, by adding states for indolent or slow growing cancers[26] and in its contribution to overdiagnosis[27]. While describing heterogeneity depends on model specification[28], it can lead to more accurate estimates but also increased uncertainty[29, 30]. The NHS-Galleri trial will not provide characterisation of within-tumour heterogeneity, so it is important to better understand its potential impacts (on detection, overdiagnosis and mortality), to support further evidence gathering in support of further model development.

Multi-cancer technologies are developing rapidly, and large and costly clinical studies are being designed and implemented across the globe. Recognising the need to produce clinical and economic evidence suitable for consideration by committees deciding whether to introduce MCED-screening programmes, it is important that similar efforts are made in the development of MCED models that make best use of the available data, and that the required data to fit those models from clinical studies is made widely available.

# Box 1 Glossary

*Stage-shift:* change in the stage distribution attributed to screening

*Sojourn time, time to transition and dwell time*: Sojourn time refers to the time spent in preclinical cancer, which is equal to the time until clinical detection or death (whichever first). Sojourn time for a particular cancer stage is the time spent in that preclinical stage of cancer, specifically it is the time until progression to the next stage, clinical detection or death (whichever first).

NHD models may be parameterised by using distributions that describe the times to each individual transition allowed in the model, for example, time for early cancer in the preclinical stage to progress to advanced preclinical cancer or time for preclinical cancer to be clinically detected.

Dwell time has been used in the literature to reflect time to stage progression, given the cancer does not get clinically detected at that stage or the individual does not die from other causes at that stage. Note that, because GRAIL models only model individuals that would be clinically diagnosed cancers under current care, the term dwell time can be used interchangeably to represent sojourn time.

*Inference*: An inferential process uses data to evaluate the NHD model and help gain understanding over the likely values of the NHD model parameters in the underlying population.

*Model identifiability*: identifiability is achieved when the number of observed quantities (the number of screen detected and interval cancers across different screens) is larger than the number of model parameters.

*Correlation in progression parameters*: Uncorrelated (or independent) parameters describing progression between stages assume that the time it takes for a cancer to progress between stage 1 and 2 is independent of the time it takes for the same cancer to progress between stages 2 and 3. These quantities may also be assumed correlated, meaning that a cancer with a lower time to progression between stage 1 and stage 2 would be expected to also present a lower time to progression in subsequent transitions. Correlation or independence can also apply to sojourn time

*Length time bias*: Length time bias is related to reliance of the models on a categorisation of disease progression across a limited number of stages. The screening models that rely on impact of screening through the stage-shift (i.e. change in the stage distribution in screening and comparator arms) ignore the possibility for additional screening benefits related to earlier cancer diagnosis not captured within allocated stages. The models that incorporate length time bias, assume that even within the same stages at diagnosis, screen-diagnosed cancers have better survival than symptomatically diagnosed cancers, due to the limitations of the stage at diagnosis as a surrogate outcome for long-term survival.

*Stage-shift assumption:* The 'stage-shift assumption' means that cases shifted to an earlier stage via screening are assumed to have the same survival as cases detected in an earlier stage without screening.

*Lead time assumption:* Lead time is defined as the time between when a cancer is detected by screening and when it would have been detected without screening. The 'lead time assumption' means mortality is not considered during lead time and therefore bringing forward diagnosis through screening does not bring forward harms such as those from more aggressive treatment.

*Cancer overdiagnosis*: We define overdiagnosis as the diagnosis, from screening, of a cancer that would not have been diagnosed under current care.

# Tables

**Table 1. List of models reviewed**

| Model | Technology | N cancers modelled | Outcomes |
|---|---|---|---|
| GRAIL models | | | |
|   Hubbell 2020 | Galleri test | 19 | Clinical |
|   Sasieni 2023 | Galleri test | 24 | Clinical |
|   Tafazzoli 2022 | Galleri test | 23 | Clinical and cost-effectiveness |
|   Dai 2024 | Galleri test | 25 | Clinical |
| Other MCED | | | |
|   Lange 2024 | Hypothetical MCED (based on Galleri test) | 12 | Clinical |
| Other multi-cancer | | | |
|   Mandrik 2024 | Dipstick test for bladder and kidney cancer | 2 | Clinical and cost-effectiveness |
|   Thomas 2024 | Imaging test for abdominal cancers | 10 | Clinical and cost-effectiveness |
| Other single-cancer | | | |
|   Skates 1991 | Blood test (CA 125) for ovarian cancer | 1 | Clinical |
|   Pinsky | CT screening for lung cancer | 1 | Clinical |

**Table 2. Key features of the Natural History of Disease (NHD) models reviewed**

| | Scope | | Structural | | | NHD model evaluation | | | Evidence for NHD | | Uncertainty |
|---|---|---|---|---|---|---|---|---|---|---|---|
| Model | Population modelled in the NHD | Mortality impacts included? | Common structure across cancer types? | Disease stages in NHD? | Clinically diagnosed cancer mortality used on screen-detected cases? | Cohort model (vs IPL) | Approach, NHD | Approach, comparative screening outcomes | Detection data on screening? | External pre-clinical progression evidence? | Uncertainty evaluated (above individual variability, where relevant)? |
| **GRAIL models** | | | | | | | | | | | |
| Hubbell 2020 | Incident | Yes | Yes | 1,2,3 and 4 | Yes | Cohort | Prediction | Prediction | No | Yes | No |
| Sasieni 2023 | Incident | Yes | Yes | 1,2,3 and 4 | Yes | Cohort | Prediction | Prediction | No | Yes | No |
| Tafazzoli 2022 | Incident | Yes | Yes | 1,2,3 and 4 | Yes | Cohort | Prediction | Prediction | No | Yes | No |
| Dai 2024 | Incident | Yes | Yes | 1,2,3 and 4 | Yes | IPL | Prediction | Prediction | No | Yes | No |
| **Other MCED** | | | | | | | | | | | |
| Lange 2024 | All | No | Yes | early vs. late | NA | Cohort | Inference, ML | Prediction | No | Yes | Yes |
| **Other multi-disease** | | | | | | | | | | | |
| Mandrik 2024 | All | Yes | No | 1,2,3 and 4 | Yes | IPL | Inference, Bayesian calibration | Prediction | No | Yes (within priors) | Yes |
| Thomas 2024 | Screen-detected | Yes | Yes | 1,2,3 and 4 | Yes | Cohort, multiple | Inference, calibration | Prediction | Yes, cases of cancer detected with screening. No data in the absence of screening | Yes | No |
| **Other single disease** | | | | | | | | | | | |
| Skates 1991 | Incident | Yes | NA | early vs. late | Yes | IPL | Prediction | Prediction | No | Yes | No |
| Pinsky | All | No | NA | early vs. late | NA | Cohort | Inference, ML | | Yes, cases of screen-detected and interval cancers by screening round | No | Yes |

Legend: NA- Not Applicable, NHD-natural history disease, IPL-Individual Patient Level; ML-Maximum Likelihood estimation

**Table 3. Assumptions over the Natural History of Disease (NHD) model**

| NHD model element | Cancer onset Distribution | Cancer onset Heterogeneity | Pre-clinical progression parameters Parameterisation | Pre-clinical progression parameters Distributions | Pre-clinical progression parameters Correlation? | Probability of clinical identification | Mortality included in NHD model | Overdiagnosis included in NHD model | Others |
|---|---|---|---|---|---|---|---|---|---|
| GRAIL models | Not modelled | | Individual parameters for progression between stages | Exponential | No | Not modelled | Yes, predictive | No | NA |
| Other MCED: Lange | Hypoexponential (fixed parameter m) | N0 | Fixed values for time in overall and late-stage pre-clinical disease | Exponential | No | Exponential | No | No | Clinical detection rates described by Poisson distribution as part of inference |
| Other multi-disease: Mandrik | Annual probability as a function of age and other risk factors (cohort model component) | Yes | Individual parameters for progression between stages, using assumptions, informative priors and constraints to ensure identification | Weibull (IPL model component) | Yes | Annual probability (cohort model component) | Yes | Yes | Bayesian calibration with multiple targets |
| Other multi-disease: Thomas | Not modelled | | Individual parameters for progression between stages | Triangular | No | Yes, for comparator arm. Triangular distribution | Yes, predictive | Yes, by considering competing mortality | NA |
| Other single disease: Skates | Not modelled | | Four-variate Normal distribution. Ratio of time in early vs late stage and constant CV for each stage were assumed constant | Four-variate Normal distribution. | Yes | Not modelled | Yes, predictive | No | NA |
| Other single disease: Pinsky | Cubic polynomial function of age | No | Single parameter for progression between early and late | Weibull (Exponential as special case) | No | Weibull (Exponential as special case) | Yes, predictive and not used in NHD model inference | Yes, predicted using NHD model estimates and external mortality estimates | NA |

Legend: NA- Not Applicable, NHD-natural history disease, IPL-Individual Patient Level; ML-Maximum Likelihood estimation

# Online appendix

**Overview of the methodology**

We performed a scoping literature review to identify previous studies that performed quantitative modelling to assess long-term clinical, epidemiological and/or economic outcomes associated with implementing MCED. Scoping searches were developed for MEDLINE and Embase in collaboration with an information specialist and are provided below. The MCED related terms were those used in a previous systematic review by Wade et. al(2024)[1] The searches aimed to identify both scientific articles and conference abstracts, with a broad remit to include statistical, economic, decision-analytic, mathematical, econometric, theoretical or epidemiological models or frameworks. The relevance of each identified record was assessed by one researcher (NK) based on its title and abstract, and if necessary, its full text. Studies were included in the broader scoping review if they performed quantitative modelling to assess long-term clinical, epidemiological and/or economic outcomes associated with implementing MCED (inclusion criteria). In this paper, we applied an additional inclusion criteria and required models to have included a Natural History of Disease component and to be available in full text.

**Search strategies**

Ovid MEDLINE(R) ALL <1946 to February 07, 2024>
08/02/2024
349 hits

1  Neoplasms/   513375
2  ((cancer$ or neoplas$ or tumour$ or tumor$ or carcinoma$ or oncolog$ or malignan$ or precancer$) adj6 (multiple$ or many or several or numerous or various or varied or miscellaneous or mixed or diverse or different)).ti,ab.   494280
3  (multicancer$ or multi-cancer$ or multitumo?r$ or multi-tumo?r$ or pan-cancer$ or pancancer$ or pan-tumo?r$ or pantumo?r$ or cross-cancer$ or crosscancer$ or cross-tumo?r$ or crosstumo?r$).ti,ab.   5384
4  ((cancer$ or neoplas$ or tumour$ or tumor$ or carcinoma$ or oncolog$ or malignan$ or precancer$) adj3 (type or types)).ti,ab.   179562
5  1 or 2 or 3 or 4   1016216
6  Liquid Biopsy/   2879
7  ((liquid$ or fluid$ or biofluid$ or bio-fluid$) adj3 biops$).ti,ab.   8961
8  6 or 7   9546
9  Biopsy/ or Biopsy, Fine-Needle/   204798
10  exp Blood/   1204680
11  9 and 10   9142
12  ((blood or h?ematolog$ or plasma or serum) adj3 biops$).ti,ab.   5680
13  11 or 12   14643
14  Hematologic Tests/   10225
15  ((blood or h?ematolog$ or plasma or serum) adj2 (test or tests or testing or tested or assay$)).ti,ab.   81266
16  14 or 15   90062
17  Multiomics/   1330
18  ((multiomic$ or multi-omic$ or panomic$ or pan-omic$ or integrative omic$) adj4 (test or tests or tested or testing or assay$ or biops$)).ti,ab.   140
19  17 or 18   1453
20  ((Multi-analyte$ or multianalyte$) adj4 (detect$ or screen$ or test or tests or tested or testing or assay$ or biops$)).ti,ab.   591

---

[1] Wade R, Nevitt S, Liu Y, Harden M, Khouja C, Raine G, Churchill R, Dias S. Multi-cancer early detection tests for general population screening: a systematic literature review. medRxiv 2024.02.14.24302576

| | | |
|---|---|---|
| 21 | 8 or 13 or 16 or 19 or 20 | 115041 |
| 22 | 5 and 21 | 6039 |
| 23 | Mass Screening/ | 117228 |
| 24 | Diagnostic Screening Programs/ | 156 |
| 25 | early diagnosis/ | 30531 |
| 26 | "Early Detection of Cancer"/ | 39411 |
| 27 | (screen$ or detect$).ti. | 672570 |
| 28 | ((early or earlystage or earli$ or first or initial or timely) adj3 (screen$ or detect$ or diagnos$ or test or tests or testing or tested)).ti,ab. | 448435 |
| 29 | (screen$ adj3 (test$ or tool$ or method$ or strateg$ or modalit$ or technolog$ or program$ or service$ or policy or policies or guideline$ or population$)).ti,ab. | 206676 |
| 30 | 23 or 24 or 25 or 26 or 27 or 28 or 29 | 1218867 |
| 31 | 22 and 30 | 1996 |
| 32 | ((cancer$ or neoplas$ or tumour$ or tumor$ or carcinoma$ or oncolog$ or malignan$ or precancer$) adj6 (multiple$ or many or several or numerous or various or varied or miscellaneous or mixed or diverse or different) adj6 (screen$ or detect$)).ti,ab. | 12389 |
| 33 | ((cancer$ or neoplas$ or tumour$ or tumor$ or carcinoma$ or oncolog$ or malignan$ or precancer$) adj6 (type or types) adj6 (screen$ or detect$)).ti,ab. | 4561 |
| 34 | 32 or 33 | 15629 |
| 35 | 21 and 34 | 635 |
| 36 | 31 or 35 | 2138 |
| 37 | (((multi-cancer$ or multicancer$ or multi-tumo?r$ or multitumo?r$) adj6 (detect$ or screen$ or test or tests or tested or testing or assay$)) or MCED or MCDBT).ti,ab. | 186 |
| 38 | ((multiple cancer$ or multiple tumo?r$) adj6 (detect$ or screen$ or test or tests or tested or testing or assay$)).ti,ab. | 552 |
| 39 | ((pan-cancer$ or pancancer$ or pan-tumo?r$ or pantumo?r$) adj6 (detect$ or screen$ or test or tests or tested or testing or assay$)).ti,ab. | 222 |
| 40 | ((cross-cancer$ or crosscancer$ or cross-tumo?r$ or crosstumo?r$) adj6 (detect$ or screen$ or test or tests or tested or testing or assay$)).ti,ab. | 5 |
| 41 | ((multi-class cancer$ or multiclass cancer$ or multi-class tumo?r$ or multiclass tumo?$) adj6 (detect$ or screen$ or test or tests or tested or testing or assay$)).ti,ab. | 6 |
| 42 | 37 or 38 or 39 or 40 or 41 | 944 |
| 43 | (Galleri or GalleriTM).mp. | 13 |
| 44 | PanSEER$.mp. | 3 |
| 45 | CancerSEEK$.mp. | 10 |
| 46 | CancerEMC$.mp. | 1 |
| 47 | (PanTum or PanTumDetect).mp. | 3 |
| 48 | Epitope-detection in monocytes.mp. | 12 |
| 49 | CancerRadar$.mp. | 0 |
| 50 | (IvyGene$ or IvyGeneCORE$).mp. | 0 |
| 51 | CancerLocator$.mp. | 1 |
| 52 | CancerDetector$.mp. | 3 |
| 53 | (EpiPanGI Dx$ or EpiPanGIDx$).mp. | 1 |
| 54 | OverC.mp. | 2 |
| 55 | DEEPGEN.mp. | 6 |
| 56 | Dxcover$.mp. | 2 |
| 57 | trucheck$.mp. | 0 |
| 58 | Elypta$.mp. | 0 |
| 59 | MiRXES$.mp. | 6 |
| 60 | Freenome$.mp. | 1 |
| 61 | 43 or 44 or 45 or 46 or 47 or 48 or 49 or 50 or 51 or 52 or 53 or 54 or 55 or 56 or 57 or 58 or 59 or 60 | 58 |
| 62 | DELFI$.mp. | 702 |
| 63 | Omni1$.mp. | 24 |
| 64 | Signal-X$.mp. | 50 |
| 65 | Harbinger$.mp. | 2137 |

66      EDIM$.mp.       182
67      LUNAR$.mp.      4624
68      MERCURY$.mp.        56259
69      62 or 63 or 64 or 65 or 66 or 67 or 68 63932
70      22 and 69       5
71      36 or 42 or 61 or 70    3023
72      exp animals/ not humans.sh.     5193858
73      71 not 72       2990
74      limit 73 to yr="2010 -Current" 2466
75      models, statistical/ or models, economic/ or models, econometric/ or models, theoretical/ or epidemiological models/       274003
76      model$.ti,ab.   3901991
77      simulation$.ti,ab.      477142
78      ((analy$ or math$ or quantitative$) adj3 framework$).ti,ab.     25303
79      (math$ adj3 (equation$ or expression$ or formula$)).ti,ab.      6805
80      75 or 76 or 77 or 78 or 79      4259368
81      74 and 80       346
82      decision support techniques/    22604
83      exp Decision Trees/     12603
84      Decision Theory/        968
85      markov chains/16081
86      (decision$ adj2 (tree$ or analy$)).ti,ab.33491
87      markov$.ti,ab.  32632
88      discrete event simulation$.ti,ab.958
89      microsimulation$.ti,ab.1979
90      82 or 83 or 84 or 85 or 86 or 87 or 88 or 89    96381
91      74 and 90       21
92      81 or 91        349

Ovid Embase <1974 to 2024 February 07>
08/02/2024
226 hits

1       neoplasm/       448117
2       ((cancer$ or neoplas$ or tumour$ or tumor$ or carcinoma$ or oncolog$ or malignan$ or precancer$) adj6 (multiple$ or many or several or numerous or various or varied or miscellaneous or mixed or diverse or different)).ti,ab.   697783
3       (multicancer$ or multi-cancer$ or multitumo?r$ or multi-tumo?r$ or pan-cancer$ or pancancer$ or pan-tumo?r$ or pantumo?r$ or cross-cancer$ or crosscancer$ or cross-tumo?r$ or crosstumo?r$).ti,ab.     8439
4       ((cancer$ or neoplas$ or tumour$ or tumor$ or carcinoma$ or oncolog$ or malignan$ or precancer$) adj3 (type or types)).ti,ab.262279
5       1 or 2 or 3 or 4 1196309
6       liquid biopsy/  12247
7       ((liquid$ or fluid$ or biofluid$ or bio-fluid$) adj3 biops$).ti,ab.15145
8       6 or 7   18095
9       biopsy/ 179967
10      exp blood/      2617042
11      9 and 10        30021
12      ((blood or h?ematolog$ or plasma or serum) adj3 biops$).ti,ab.10814
13      11 or 12        39507
14      blood examination/      19256
15      ((blood or h?ematolog$ or plasma or serum) adj2 (test or tests or testing or tested or assay$)).ti,ab.   131094
16      14 or 15        146604

| | | |
|---|---|---|
| 17 | ((multiomic$ or multi-omic$ or panomic$ or pan-omic$ or integrative omic$) adj4 (test or tests or tested or testing or assay$ or biops$)).ti,ab. | 261 |
| 18 | ((Multi-analyte$ or multianalyte$) adj4 (detect$ or screen$ or test or tests or tested or testing or assay$ or biops$)).ti,ab. | 796 |
| 19 | 8 or 13 or 16 or 17 or 18 | 201045 |
| 20 | 5 and 19 | 14830 |
| 21 | mass screening/ | 62609 |
| 22 | cancer screening/ | 100611 |
| 23 | early cancer diagnosis/ | 14712 |
| 24 | (screen$ or detect$).ti. | 812818 |
| 25 | ((early or earlystage or earli$ or first or initial or timely) adj3 (screen$ or detect$ or diagnos$ or test or tests or testing or tested)).ti,ab. | 660791 |
| 26 | (screen$ adj3 (test$ or tool$ or method$ or strateg$ or modalit$ or technolog$ or program$ or service$ or policy or policies or guideline$ or population$)).ti,ab. | 303302 |
| 27 | 22 or 23 or 24 or 25 or 26 | 1579016 |
| 28 | 20 and 27 | 4400 |
| 29 | ((cancer$ or neoplas$ or tumour$ or tumor$ or carcinoma$ or oncolog$ or malignan$ or precancer$) adj6 (multiple$ or many or several or numerous or various or varied or miscellaneous or mixed or diverse or different) adj6 (screen$ or detect$)).ti,ab. | 17884 |
| 30 | ((cancer$ or neoplas$ or tumour$ or tumor$ or carcinoma$ or oncolog$ or malignan$ or precancer$) adj6 (type or types) adj6 (screen$ or detect$)).ti,ab. | 6616 |
| 31 | 29 or 30 | 22608 |
| 32 | 19 and 31 | 1136 |
| 33 | 28 or 32 | 4698 |
| 34 | (((multi-cancer$ or multicancer$ or multi-tumo?r$ or multitumo?r$) adj6 (detect$ or screen$ or test or tests or tested or testing or assay$)) or MCED or MCDBT).ti,ab. | 400 |
| 35 | ((multiple cancer$ or multiple tumo?r$) adj6 (detect$ or screen$ or test or tests or tested or testing or assay$)).ti,ab. | 899 |
| 36 | ((pan-cancer$ or pancancer$ or pan-tumo?r$ or pantumo?r$) adj6 (detect$ or screen$ or test or tests or tested or testing or assay$)).ti,ab. | 515 |
| 37 | ((cross-cancer$ or crosscancer$ or cross-tumo?r$ or crosstumo?r$) adj6 (detect$ or screen$ or test or tests or tested or testing or assay$)).ti,ab. | 10 |
| 38 | ((multi-class cancer$ or multiclass cancer$ or multi-class tumo?r$ or multiclass tumo?$) adj6 (detect$ or screen$ or test or tests or tested or testing or assay$)).ti,ab. | 11 |
| 39 | 34 or 35 or 36 or 37 or 38 | 1743 |
| 40 | (Galleri or GalleriTM).mp. | 47 |
| 41 | PanSEER$.mp. | 8 |
| 42 | CancerSEEK$.mp. | 20 |
| 43 | CancerEMC$.mp. | 1 |
| 44 | (PanTum or PanTumDetect).mp. | 7 |
| 45 | Epitope-detection in monocytes.mp. | 18 |
| 46 | CancerRadar$.mp. | 1 |
| 47 | (IvyGene$ or IvyGeneCORE$).mp. | 9 |
| 48 | CancerLocator$.mp. | 1 |
| 49 | CancerDetector$.mp. | 2 |
| 50 | (EpiPanGI Dx$ or EpiPanGIDx$).mp. | 2 |
| 51 | OverC.mp. | 1 |
| 52 | DEEPGEN.mp. | 13 |
| 53 | Dxcover$.mp. | 13 |
| 54 | trucheck$.mp. | 4 |
| 55 | Elypta$.mp. | 1 |
| 56 | MiRXES$.mp. | 53 |
| 57 | Freenome$.mp. | 84 |
| 58 | 40 or 41 or 42 or 43 or 44 or 45 or 46 or 47 or 48 or 49 or 50 or 51 or 52 or 53 or 54 or 55 or 56 or 57 | 269 |
| 59 | DELFI$.mp. | 1285 |
| 60 | Omni1$.mp. | 135 |
| 61 | Signal-X$.mp. | 1059 |
| 62 | Harbinger$.mp. | 3074 |
| 63 | EDIM$.mp. | 268 |
| 64 | LUNAR$.mp. | 8449 |
| 65 | MERCURY$.mp. | 73242 |
| 66 | 59 or 60 or 61 or 62 or 63 or 64 or 65 | 87469 |
| 67 | 20 and 66 | 25 |
| 68 | 33 or 39 or 58 or 67 | 6420 |
| 69 | (animal/ or animal experiment/ or animal model/ or animal tissue/ or nonhuman/) not exp human/ | 6917818 |
| 70 | 68 not 69 | 6305 |
| 71 | limit 70 to yr="2010 -Current" | 5690 |
| 72 | exp economic model/ | 4068 |
| 73 | statistical model/ | 176506 |
| 74 | mathematical model/ | 141694 |
| 75 | epidemiological model/ | 691 |
| 76 | model$.ti,kw. | 963001 |
| 77 | ((economic or econometric or cost$ or statistical or mathematical or epidemiological or state transition or interception or natural history) adj2 model$).ab. | 123461 |
| 78 | simulation$.ti,ab. | 487641 |
| 79 | ((analy$ or math$ or quantitative$) adj3 framework$).ti,ab. | 28327 |
| 80 | (math$ adj3 (equation$ or expression$ or formula$)).ti,ab. | 7985 |
| 81 | or/72-80 | 1662473 |
| 82 | 71 and 81 | 192 |
| 83 | "decision tree"/ | 23369 |
| 84 | decision theory/ | 1861 |
| 85 | exp markov chain/ | 16817 |
| 86 | (decision$ adj2 (tree$ or analy$ or model$)).ti,ab. | 53926 |
| 87 | markov$.ti,ab. | 40975 |
| 88 | discrete event simulation/ | 439 |
| 89 | discrete event simulation$.ti,ab. | 1478 |
| 90 | microsimulation$.ti,ab. | 2914 |
| 91 | or/83-90 | 107658 |
| 92 | 71 and 91 | 56 |
| 93 | 82 or 92 | 226 |

**Summary of results**

The search strategy was run on February 8, 2024, and retrieved 575 records in total, with 480 unique records remaining after duplicates were removed. The searches had an objective that was broader than the review in this paper. They aimed to scope the modelling literature, and presented a broad remit which included statistical, economic, decision-analytic, mathematical, econometric, theoretical or epidemiological models or frameworks. In this paper, we have only looked at models with an explicit natural history of disease (NHD) component.

For the broader scoping review 20 publications/abstracts were selected. These were split into clinical models or clinical and cost-effectiveness models. Ten were clinical models, five of which were full papers and another five were abstracts. The remaining 10 included cost-effectiveness, 3 of which were full papers and the remaining seven were abstracts. Where the work in an abstract used a particular model published in full text elsewhere, the abstract was linked to the full model publication. Given information was often limited, we confirmed the attribution of abstracts to models with GRAIL.

Only five of the above MCED models had an explicit NHD component, and therefore only these were included in the review in this paper. Four of these were funded by GRAIL and specifically related to the Galleri test [13-16] and one was based on a hypothetical MCED (although using some data inputs derived from studies on the Galleri test) [17].